\documentclass[runningheads]{llncs}
\usepackage{graphicx}
\usepackage{hyperref}
\usepackage{xcolor}
\usepackage{booktabs}
\newcommand\blfootnote[1]{%
	\begingroup
	\renewcommand\thefootnote{}\footnote{#1}%
	\addtocounter{footnote}{-1}%
	\endgroup
}
\usepackage{amsmath} 
\usepackage{hyphenat}

\begin{document}
\title{Improved prosodic clustering for multispeaker and speaker-independent phoneme-level prosody control}
%
\titlerunning{Improved clustering for speaker-independent phoneme-level prosody control}
%
\author{Myrsini Christidou\inst{1}\thanks{Equal contribution} \and
	Alexandra Vioni\inst{1}$^{\star}$ \and
	Nikolaos Ellinas\inst{1} \and
	Georgios Vamvoukakis\inst{1} \and
	Konstantinos Markopoulos\inst{1} \and
	Panos Kakoulidis\inst{1} \and
	June Sig Sung\inst{2} \and
	Hyoungmin Park\inst{2} \and
	Aimilios Chalamandaris\inst{1} \and
	Pirros Tsiakoulis\inst{1}}
\authorrunning{M. Christidou et al.}
%
\institute{Innoetics, Samsung Electronics, Greece \and
	Mobile Communications Business, Samsung Electronics, Republic of Korea\\	
	\email{\{m.christidou, a.vioni\}@partner.samsung.com, \\
		\{n.ellinas, g.vamvouk\}@samsung.com }}

\maketitle              
\begin{abstract}
This paper presents a method for phoneme-level prosody control of F0 and duration on a multispeaker text-to-speech setup, which is based on prosodic clustering.
An autoregressive attention-based model is used, incorporating multispeaker architecture modules in parallel to a prosody encoder.
Several improvements over the basic single-speaker method are proposed that increase the prosodic control range and coverage.
More specifically we employ data augmentation, F0 normalization, balanced clustering for duration, and speaker-independent prosodic clustering.
These modifications enable fine-grained phoneme-level prosody control for all speakers contained in the training set, while maintaining the speaker identity.
The model is also fine-tuned to unseen speakers with limited amounts of data and it is shown to maintain its prosody control capabilities, verifying that the speaker-independent prosodic clustering is effective.
Experimental results verify that the model maintains high output speech quality and that the proposed method allows efficient prosody control within each speaker's range despite the variability that a multispeaker setting introduces.

\end{abstract}
\section{Introduction}

\blfootnote{The final authenticated publication is available online at \url{https://doi.org/10.1007/978-3-030-87802-3_11}}

Neural text-to-speech (TTS) systems, such as Tacotron \cite{wang2017tacotron,shen2018natural}, that produce synthetic voice of high quality and naturalness, have paved the way for exploration and control over more elaborate aspects of speech, including prosody and style.
In the recent surge of multispeaker and multilingual architectures \cite{ping2018deep,zhang2019learning}, it is valuable to integrate prosodic control mechanisms in such systems by taking advantage of the speaker and prosodic diversity, while also embracing and overcoming the challenges that arise from this heterogeneity.

\subsection{Related Work}

Since basic neural TTS systems model the average speaking style of the training data, style control is traditionally exerted by using extended models conditioned on a style embedding space \cite{Skerry-Ryan2018} that is learned in an unsupervised way.
These approaches introduce a Global Style Tokens (GSTs) module \cite{GSTs}, or a Variational Autoencoder (VAE) \cite{hsu2018hierarchical,battenberg2019effective,Zhang2019} to learn latent representations of prosody in utterance-level prosody control or transfer.
In addition, fine-grained control at frame-level, phoneme-level or word-level resolutions can be achieved using temporal constraints \cite{Lee2019}, hierarchical modeling \cite{Sun2020,Sun2020a,chien2020hierarchical}, or adversarial learning \cite{daxin2020finegrained}.
In \cite{du2021mixture}, phoneme-level prosody embeddings are modeled using a Gaussian mixture model (GMM) based mixture density network (MDN).

Explicit control over prosodic features such as F0 and duration by extracting these features and using them as input to a prosody encoder has been implemented for utterance-level \cite{Shechtman2019,Gururani2019,Raitio2020} and more fine-grained \cite{Wan2019,Park2019,Klimkov2019,Zhang2020} control.
In \cite{angelini2020singing}, data augmentation is applied to extend the voice range in terms of F0 and duration, and note embeddings are used in parallel to the phoneme sequence, to pursue singing synthesis.
\cite{KiyoshiKURIHARA20212020EDP7104} inserts prosodic symbols to the phoneme sequence to model accents, pauses, and sentence endings.

Regarding multispeaker systems, several approaches for prosody control and transfer have been carried through.
Mellotron \cite{valle2020mellotron} constitutes a multispeaker voice synthesis model conditioned on rhythm and continuous pitch contours, that can generate expressive and singing voice.
In \cite{neekhara2021expressive}, a controllable voice cloning model based on Tacotron 2, conditioned on a speaker embedding, pitch contour and latent style tokens, enables fine-grained style transfer for an unseen speaker.
\cite{kumar2020shot} proposes a method for few-shot speaker adaptation and generation of an unseen speaker's style by incorporating a non-autoregressive feed-forward Transformer along with adaptive normalization.
Adversarial learning was employed in \cite{karlapati2020copycat} to avoid source speaker leakage in prosody transfer tasks, and in \cite{wang2021adversarially} to ensure prosodic disentanglement in voice conversion.
Also, in \cite{zhang2020voice}, a multispeaker Transformer-based model with an ASR module and an utterance-level prosody encoder is fine-tuned to the target speaker for prosody transfer.

However, all the aforementioned approaches for multispeaker prosody control refer to utterance level control; or to a generic prosodic style in the case of a more detailed scale.
Alternatively, we propose speaker independent phoneme-level prosody control over F0 and duration in a multispeaker system.
Furthermore, we show that the proposed method is feasible even for new speakers with very limited data, as the system maintains the prosody control capability after fine-tuning the model. 

\subsection{Proposed Method}

In this paper we apply prosodic clustering for multispeaker phoneme-level \linebreak prosody control of F0 and duration in TTS.
The discretization and grouping of these features into clusters provides great controllability over prosody in synthesized speech \cite{Vioni2021}, however limited inside the speaker's range since the outermost clusters contain more extreme values which are not frequent in the training data.
Using the model for unseen speaker adaptation, would also cause the single speaker clustering method to fail due to the different speaker characteristics.

To address these problems, we have incorporated several preprocessing steps and we have adapted the prosodic clustering method for multispeaker TTS.
We apply augmentation transformations to the training data \cite{angelini2020singing}, in order to increase the number of samples in the outermost clusters.
To minimize stability issues in duration control, we have adopted a balanced duration clustering strategy, assigning an equal number of samples in each cluster instead of using K-Means, as it was done in the single speaker model for both f0 and duration.
In addition, we introduce per speaker normalization of F0 with the purpose of neutralizing speaker and gender variations, essentially offering speaker-independent control over the same prosodic space.

The aforementioned modifications allow us to create universal, speaker\hyp{}independent clusters for F0 and duration.
Thus, prosody control capabilities can be extended to previously unseen speakers, after fine-tuning the model with only a few samples.
Overall, the contributions of this work are the following: (i) data augmentation for increased control range, (ii) speaker independent clustering via F0 normalization, (iii) balanced duration clustering for better stability, and (iv) speaker adaptation with limited data while maintaining controllability. 

\section{Method} \label{method}

\subsection{Multispeaker model} \label{multispeaker_model}

The model presented in \cite{Vioni2021} is adapted to a multispeaker architecture.
It is an autoregressive attention-based text-to-speech model, that receives an input sequence of phonemes $\boldsymbol{p}=[p_1,...,p_N]$ and sequences of F0 and duration tokens $\boldsymbol{f}=[f_1,...,f_M]$, $\boldsymbol{d}=[d_1,...,d_M]$ which are jointly referred to as prosodic features.

Each phoneme has a corresponding token for F0 and duration, while word boundaries and punctuation marks do not receive any such tokens, therefore ${M<N}$. 
The phoneme sequence is passed into a text encoder which produces a text encoder representation $\boldsymbol{e}=[e_1,...,e_N]$ and the prosodic feature sequences are concatenated and then passed into a prosody encoder which produces the prosody encoder representation $\boldsymbol{e'}=[e'_1,...,e'_M]$. 
On the decoder side, apart from the MoL (Mixture of Logistic distributions) attention module responsible for producing a robust phoneme alignment \cite{Ellinas2020}, we use an extra module for the alignment of the prosodic representations.
As mentioned in \cite{Vioni2021}, we do not wish the prosodic attention context vector to contain any phoneme information, so we choose to process the prosodic sequence with a separate attention module to enforce this condition.

\begin{figure}
	\includegraphics[width=\textwidth]{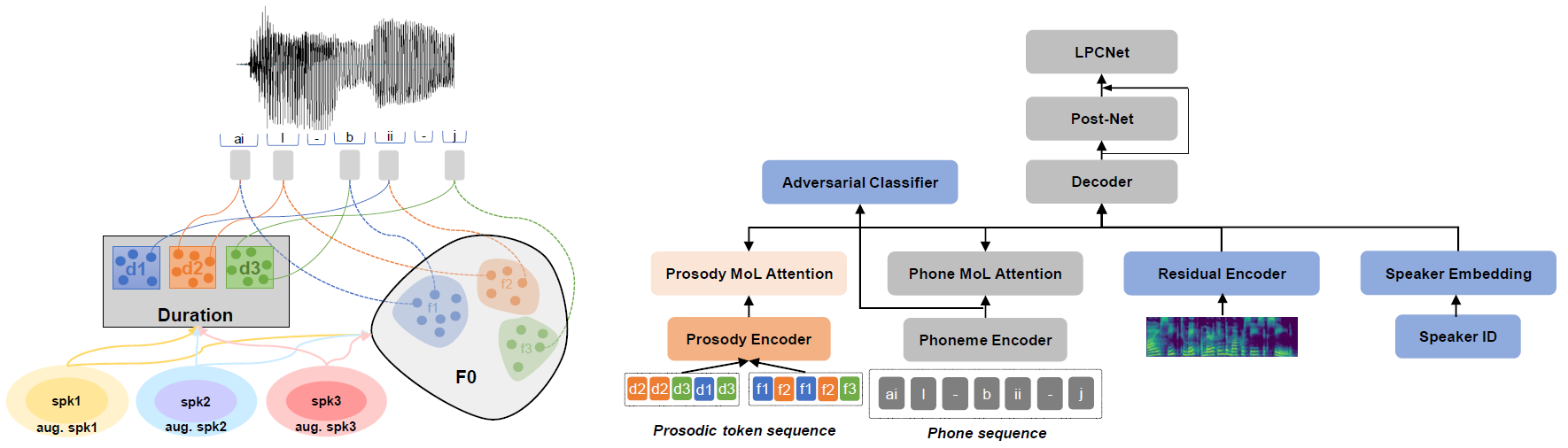}
	\caption{Multispeaker Prosody Control Model.
		The prosodic features are firstly clustered. 
		The normalization applied to each cluster is common for every speaker. 
		The single speaker architecture was modified with an Adversarial Classifier, Speaker Embedding and Residual Encoder in order to enable the model to train on multiple speakers.} \label{fig:architecture}
\end{figure}

Each speaker is mapped to a 64-dimensional learnable embedding, which is used to condition the decoder.
A variational residual encoder \cite{hsu2018hierarchical} is implemented to model any additional information included in the audio samples other than speaker identity, text and prosodic features, like acoustic conditions and noise.
An adversarial speaker classifier similar to \cite{zhang2019learning} is also added, to induce disentanglement of the phoneme representations and the speakers' identity.
The architecture is presented in Figure \ref{fig:architecture}.

\vspace{12pt}

\subsection{Dataset augmentation} \label{dataset_augmentation}

In order to make our system more robust, we have employed voice data augmentation to widen the speaker range regarding F0 and duration, and to increase the number of samples in each cluster, as seen in previous multispeaker \cite{Cooper2020} and singing synthesis \cite{blaauw2017,angelini2020singing} papers.
The data transformations applied are: pitch shifting by [-6, -4, -2, 2, 4, 6] semitones, and tempo changes by altering speaking rate to [0.70, 0.80, 0.90, 1.10, 1.20, 1.30] of the original one, using the Praat Vocal Toolkit \cite{corretge2012pvt}.
The selection of alterations to be applied on the utterances was made similarly to \cite{angelini2020singing} so that adequate augmentation was added to enhance the training data, whilst no distortion was introduced to the produced audio.
Specifically, we have applied one of these twelve transformations to each spoken utterance in our dataset, with the resulting dataset being double in size compared to the initial one.

We perform clustering on our augmented dataset together with the original one to get new cluster centroids that correspond to the widened F0 and duration ranges.
The proposed augmentation method extends the prosodic range, while also enhancing model robustness and voice quality.

\subsection{F0 normalization}

Since different speakers have different pitch ranges it is not practical to cluster the F0 values of each speaker separately.
We first apply $z$-score normalization to each speaker's extracted F0 values, and then cluster the normalized values of all speakers together to obtain universal F0 centroids. Thus for each phoneme's corresponding pitch in the dataset we apply: 

\vspace{10pt}
\begin{equation}
	F_0 =\frac{f-\mu_i}{\sigma_i} , i \in speakers
\end{equation}
\vspace{5pt}

\noindent where $f$ is the unnormalized F0 and $\mu$, $\sigma$ are the mean and variance of the respective speaker's F0 values.
This way, we deal with gender and speaker variation in pitch and create a mapping from each speaker's F0 values to a common prosodic space, where clustering can be performed universally.
The normalization method also facilitates adding new speakers, because the normalized F0 values of the new speakers can be directly mapped to the universal centroids without the need of recomputing them.

\subsection{Balanced duration clustering}

Regarding phoneme durations, it is observed that, for the same phone, typical duration ranges are similar across all speakers, thus duration normalization is not necessary before clustering.
However, results from our previous work \cite{Vioni2021} show that voice quality deterioration is mostly noted in duration control when using the outermost clusters, and not so much in F0 control.
Thus, we have adopted a different scheme for extracting duration clusters.
Following the single speaker method, clustering is still performed per phoneme, but the K-means algorithm is replaced with a simple balanced clustering method; the average phoneme duration values are sorted in ascending order and grouped into the desired number of intervals, so that each interval contains an equal number of samples.
We observe that using this grouping strategy slightly decreases the duration control range, as extreme values are averaged out by being pushed towards the bulk of more frequent phoneme durations, but more importantly increases the duration control stability.

\subsection{Speaker adaptation}

The same method as described in the sections above was used to investigate the feasibility of fine-grained prosody control on a previously unseen speaker with a very small number of samples.
Extra attention was given so that the selected sentences, that were to be used in the new training, would provide enough phonetic coverage. 
This means that the utterances would contain each phoneme at least once. 

After applying augmentation and $z$-score normalization to the new speaker's data, we fine-tuned our pretrained model by replacing one of the speakers in the training set with the new speaker.
We experimented with various recording time lengths in order to test the model's limits and investigate how many minutes of recorded speech is needed to achieve similar quality results with the speakers in the training set.
We found that even with as few as 5.7 minutes of recordings our model was able not only to reproduce that speaker's voice, but also to manipulate phoneme-level F0 and duration in a similar manner to the voices in our training set.

\section{Experiments and Results}

An initial multispeaker model is trained on an internal dataset containing three female and two male speakers for a total of 159.7 hours of speech.
The preprocessing methods described in section \ref{method} are applied to all voice data.
A randomly selected augmentation is applied to each utterance in order to create new augmented speakers, doubling the size of the original multispeaker dataset.
Considering each speaker together with their augmented version as a new set, $z$-score normalization of the extracted F0 values is applied for each speaker separately.
The duration labels are computed with balanced clustering, while the K-Means algorithm was used to find the optimal centroids for F0 values, with the selected number of clusters both for F0 and duration being fixed to 15.
Since the duration values do not vary much from speaker to speaker and the F0 values are normalized, the values of each feature lie in the same space, so the two methods are applied on the full augmented dataset with all of the speakers mixed together.

In order to evaluate the effectiveness of adapting the multispeaker model to unseen speakers, we use the voice of Catherine Byers (Cathy) from the 2013 Blizzard Challenge in two training setups.
The first one, referred to as \textit{Cathy-multi}, is a model trained on the multispeaker dataset mentioned above, with the addition of Cathy, resulting in a total duration of 223.9 hours.
This model is used to generate samples from the target speaker when they are included in the initial training, utilizing their full set of data.
For the second model, \textit{Cathy-adapt}, we select 100 recordings from the Cathy dataset containing 7.72 minutes of speech and fine-tune the initial 5-voice multispeaker model.
The selection of recordings is done with a method introduced in \cite{Chalamandaris2009} which sorts utterances of a speech corpus in descending order of phonological diversity, in order to maximize the phonetic coverage in a small collection of recordings.

The initial multispeaker model is also adapted to another 2 female and 2 male unseen voices by applying the speaker adaptation process independently for each one, in order to obtain results from different target speakers and genders.
For this task, we use the LJ Speech dataset \cite{ljspeech17}, an audiobook male voice and two additional internal voices, one female and one male.
The corpus selection process to ensure phonetic coverage with 100 sentences for each voice resulted in 10.24, 5.7, 10.83, and 13.17 minutes of speech respectively.
By using these limited data we obtain the respective speaker adapted models, namely \textit{LJ-adapt}, \textit{Audiobook-male-adapt}, \textit{Female-adapt} and \textit{Male-adapt}.

The same preprocessing steps for augmentation, duration and F0 clustering as described above are applied to the adaptation speakers.
An important change is that, since the goal is to integrate the new speaker into the already trained model, the duration intervals and F0 centroids are not recomputed, but rather used as pretrained values from the full length multispeaker dataset in order to find the corresponding target speaker values.
This process does not introduce any inconsistencies as the normalized F0 values of the new speakers lie in the same universal space with the initial training dataset.
The model is fine-tuned for 5K iterations as a single speaker model, after replacing one of the initial speaker identities with the target speaker to obtain the desired voice characteristics.

The voices used in our experiments were of native US English speakers.
All audio data was resampled at 24 kHz and the extracted acoustic features consist of 20 Bark-scale cepstral coefficients, the pitch period and pitch correlation, in order to match the modified LPCNet vocoder \cite{Vipperla2020}. The proposed model follows the same architecture as in \cite{Vioni2021} for the phoneme encoder, prosody encoder, attention mechanism and decoder with the additions described in \ref{multispeaker_model}.

For the objective and subjective tests we selected 100 utterances from Catherine Byers' dataset, which were excluded from the training data. Those were used to extract the ground truth prosodic labels. In general, the model is able to synthesize arbitrary text provided that the corresponding prosodic labels are  specified; predicted by a separate model, or extracted from a reference utterance.


\vspace{5pt}
\subsection{Objective evaluation}
\vspace{5pt}

In order to evaluate the control capability of the proposed model, a test set is generated for each speaker by assigning the prosody tokens of each sentence to a single cluster in an ascending order.
Specifically, this process is applied at one prosodic category at a time, keeping the other category's tokens at their ground truth values.

\begin{figure}
	\includegraphics[width=\textwidth]{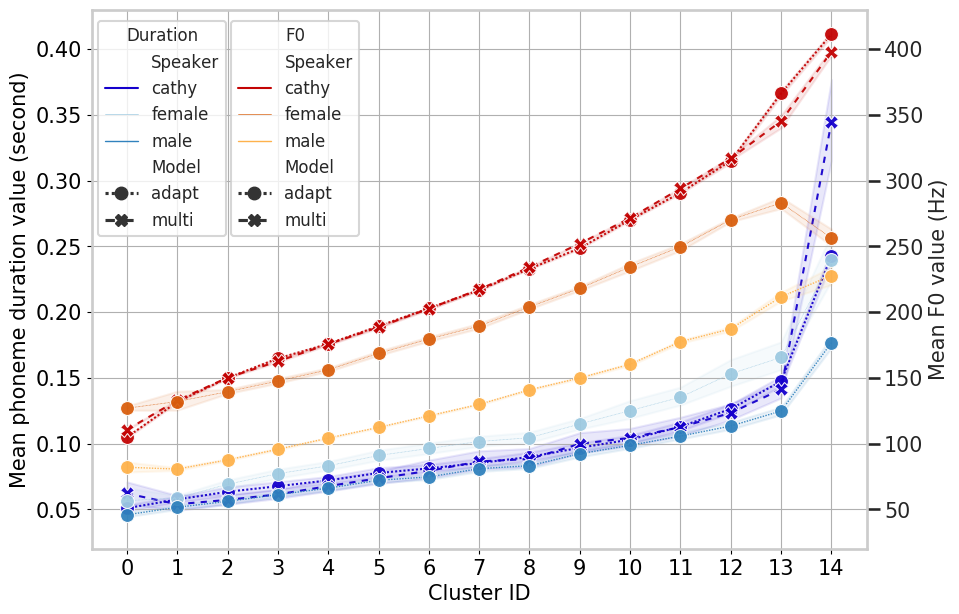}
	\caption{Sentence level mean F0 and average phoneme duration for ascending cluster IDs with 95\% confidence intervals.
		The left y-axis corresponds to the duration graphs while the right y-axis corresponds to the F0 graphs.} \label{fig:ascending}
\end{figure}

In Figure \ref{fig:ascending} the mean values of F0 and phoneme duration are depicted, calculated over the extracted features of every synthesized test utterance modified according to the specific cluster ID shown in the horizontal axis.
The depicted models belong to the configurations Cathy-multi, Cathy-adapt, Female-adapt and Male-adapt.
We can observe that all models follow the ascending order of the cluster IDs both in F0 and duration variations, proving that controllability is retained in the multispeaker setup.
Cathy-multi and Cathy-adapt perform alike and obtain similar values for the same cluster IDs, proving that prosody control in the same range is possible even with a few data, compared to a large speaker dataset.
The rest of the adaptation models present a same ascending behavior, with the male voices assuming lower F0 values from the female ones, despite being trained with a common set of prosodic labels, indicating that our method is indeed speaker and gender independent.

Audio samples of utterance-level and word-level prosodic control are available on our website.
Phoneme-level manipulation of F0 and duration is also demonstrated, for which no established protocol exists for objective evaluation.
We encourage readers to listen to the audio samples: \url{https://innoetics.github.io/publications/multispeaker-prosody-control/index.html}

\subsection{Subjective evaluation}

We performed listening tests in order to assess the quality of the proposed method, with respect to naturalness and speaker similarity.
Regarding naturalness, the set of the chosen 100 sentences which were excluded from the training data and modified in terms of F0 and duration were used to synthesize voice samples from Cathy-multi, Cathy-adapt and Male-adapt models.
Listeners were asked to score the samples' naturalness on a 5-point Likert scale.

F0 and duration modification was done by adding or subtracting an offset from the ground truth prosodic labels of each test sentence, with the offsets varying in range $[-11,+11]$.
This was straightforward, since the proposed labels for both prosodic features range from 0-14 and are represented as so.
Regarding F0, adding or subtracting an offset from the ground truth value leads to synthesized voice with higher or lower pitch, whilst regarding duration, these offsets lead to slower or faster uttered phones, respectively.
This method, which verifies the controllability of the model, is now used to evaluate naturalness.

\begin{figure}
	\includegraphics[width=\textwidth]{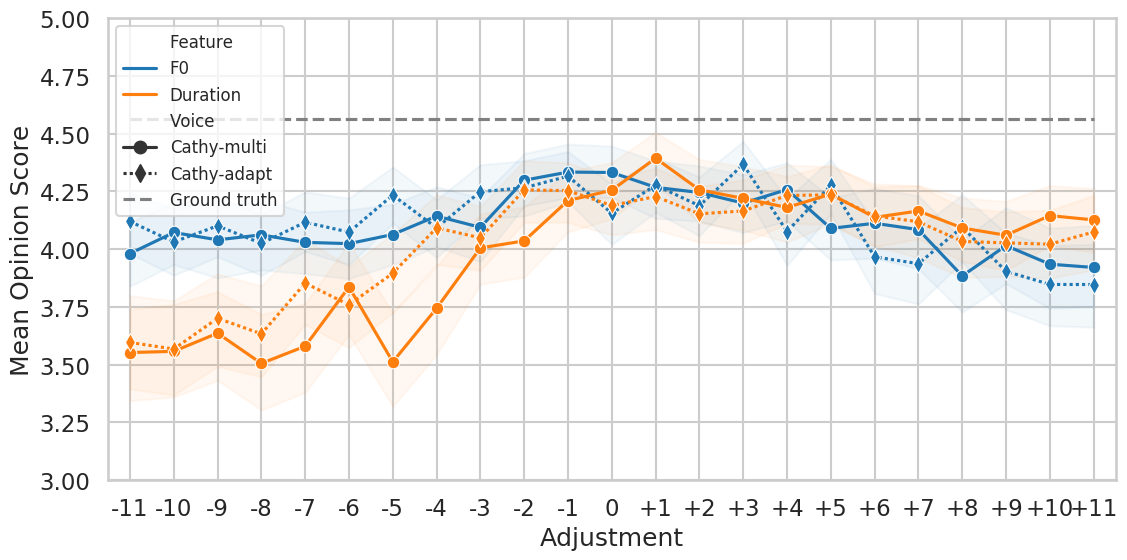}
	\caption{Mean opinion scores for Cathy-multi and Cathy-adapt with 95\% confidence intervals} \label{fig:mos-cathy}
\end{figure}


\begin{figure}
	\includegraphics[width=\textwidth]{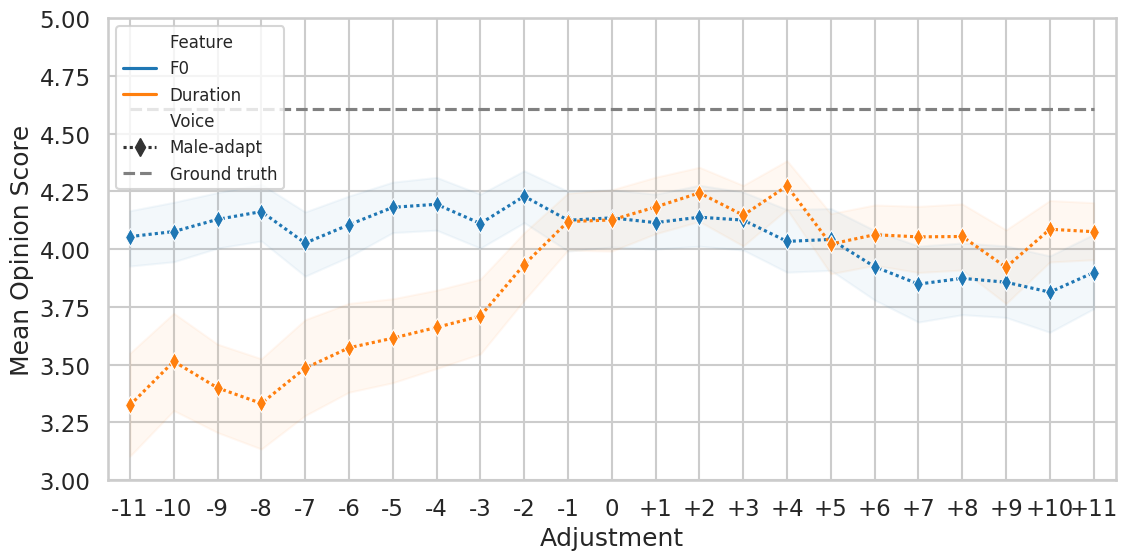}
	\caption{Mean opinion scores for Male-adapt with 95\% confidence intervals} 
	\label{fig:mos-jsj}
\end{figure}

In order to facilitate the Mean Opinion Score (MOS) results' visualization, each prosodic feature was modified independently, while the labels of the other feature retained their ground truth values.
In total, 2400 test utterances were rated for naturalness, with each one receiving 20 scores by native speakers via the Amazon Mechanical Turk.

The MOS is depicted as a function of the modification offset in Figure \ref{fig:mos-cathy} for the Cathy-multi and the Cathy-adapt models, and in Figure \ref{fig:mos-jsj} for the Male-adapt model.
Based on the plots, it can be said that the voice samples with modified prosodic tokens retain reasonable naturalness levels in general, with the exception of very low duration offsets.
These offsets correspond to extremely fast speech which is generally considered unnatural.
Moreover, MOS scores of the voice samples produced by Cathy-adapt are directly comparable in naturalness with the scores of Cathy-multi, over the full modification range, as it can be seen in Figure \ref{fig:mos-cathy}.
Hence, it is shown that, despite been trained with very limited data, the speaker-adapted models' capability of prosodic modification also preserves high voice naturalness, in levels similar to the multispeaker model, which was trained with the full dataset.

Regarding speaker similarity, listening tests were performed to evaluate \linebreak speaker adaptation with limited data.
For each speaker adapted model, 20 samples synthesized with ground truth prosodic labels were compared to a reference audio of the respective speaker.
Listeners were asked to rate speaker similarity on a 5-point Likert scale.
Each utterance received 40 scores by native speakers via the Amazon Mechanical Turk.


\begin{table}
	\caption{Speaker similarity MOS for speaker adaptation with 95\% confidence intervals}
	\label{tab:spksim}
	\vspace{20pt}
	\centering
	\begin{tabular}{@{\hspace*{2mm}} l @{\hspace*{4mm}} l @{\hspace*{2mm}}}
		\toprule
		\textbf{Voice} & \textbf{Speaker Similarity} \\
		\midrule
		Cathy-adapt & $3.506 \pm 0.140$ \\
		LJ-adapt & $3.000 \pm 0.142$ \\
		Female-adapt & $3.858 \pm 0.124$\\
		Male-adapt & $3.942 \pm 0.117$ \\
		Audiobook-male-adapt & $3.633 \pm 0.121$ \\
		\bottomrule
	\end{tabular}
\end{table}

By observing the results in Table \ref{tab:spksim}, it is evident that speaker similarity is adequate for all voices, taking into consideration that speaker adaptation was performed with only few minutes of speech from each speaker, and very satisfactory for the internal voices.
Our speaker similarity MOS scores range between values in the same area as the MOS of previous works on multispeaker and speaker adaptation TTS \cite{zhang2020voice,zhang2019learning}.
It can be said that the speaker similarity MOS scores correlate well with the voice recordings quality, since they are higher for internal voices with clear recordings, but deteriorate for voice datasets where noise and artifacts are present.

\section{Conclusions}

In this paper, we expand upon our previous work on fine-grained prosody control from a single speaker to any number of speakers without compromising the quality of generated speech. 
We apply augmentation, feature normalization and a universal clustering method for all speakers' recordings so that we can produce universal F0 and duration clusters for training.
The same principles are applied to new, previously unseen speakers with very few recordings, in order to test if this method can be used to create synthetic speech similar to the target voice with the same quality and level of control.

Our experiments verify that the multispeaker and speaker adapted models retain the control capability over F0 and duration and generate high quality speech, independently of gender or different voice characteristics.
Moreover, the speaker adapted models' scores indicate reasonable similarity to the original speakers' audio, given the short duration and variable quality of the recordings across speakers.



%
%
%
\bibliographystyle{splncs04}
\bibliography{mybib}

\end{document}